# Heterogeneously integrated silicon photonics for the mid-infrared and spectroscopic sensing

Yu Chen[1*], Hongtao Lin[2*], Juejun Hu[2], and Mo Li[1†]

[1]Department of Electrical and Computer Engineering, University of Minnesota, Minneapolis, MN 55455, USA

[2] Department of Material Science and Engineering, University of Delaware, Newark, DE 19716, USA

ABSTRACT

**Besides being the foundational material for microelectronics, in optics, crystalline silicon has long been used for making infrared lenses and mirrors. More recently, silicon has become the key material to achieve large-scale integration of photonic devices for optical interconnect and telecommunication[1, 2]. For optics, silicon has significant advantages: it offers a very high refractive index and is highly transparent in the spectral range from 1.2 to 8 μm. To fully exploit silicon's superior performance in a remarkably broad range and to enable new optoelectronic functionalities, here we describe a general method to integrate silicon photonic devices on arbitrary substrates. In particular, we apply the technique to successfully integrate silicon micro-ring resonators on mid-infrared compatible substrates for operation in the mid-infrared[3, 4, 5, 6, 7]. These high-performance mid-infrared optical resonators are utilized to demonstrate, for the first time, on-chip cavity-enhanced mid-infrared spectroscopic analysis of organic chemicals with a limit of detection of less than 0.1 ng.**

The traditional silicon-on-insulator (SOI) platform provides a very high index contrast between silicon ($n$ = 3.48) and its cladding, silicon dioxide ($n$ = 1.45), enabling low-loss, ultrahigh density integration of silicon photonics in the near-infrared (near-IR) telecommunication band[8]. However, the use of silicon oxide as the cladding also imposes a limitation on the spectral range that silicon photonics can operate within. Although silicon is transparent in the ~1−8 μm wavelength range[3, 4, 9], silicon dioxide starts to absorb light strongly above ~4 μm. Thus, a very wide wavelength range from ~4 to 8 μm where remarkable performance can be obtained with silicon becomes inaccessible to the SOI platform. In addition, since amorphous silicon dioxide has no electro-optical effects and only exhibits very weak nonlinearity, the oxide layer in SOI merely acts as an inert cladding material providing no other functionalities much needed for integrated photonic systems, such as on-chip light modulation and nonlinear optical signal processing. If integrated silicon photonic devices can be built on materials that are transparent over a wider band, have strong electro-optical and nonlinear optical effects, or possess optical gain, the potential of silicon photonics with advanced active photonic functionalities, in addition to their established role as a passive optical platform, can be fully leveraged in future large scale integrated photonic systems. However, to date, direct growth or bonding of crystalline silicon on materials other than silicon dioxide remains to be difficult, unreliable or too expensive for large scale integrated device[10, 11, 12].

---

[*] These authors contributed equally to this work.

[†] Corresponding author: moli@umn.edu

Here we describe a method to unleash silicon's tremendous potential for integrated photonics by transferring silicon photonic devices from the traditional SOI platform to new substrates. Notably this method does not involve the use of adhesives that may compromise the pristine optical properties of silicon, especially in the infrared, and induce process complexity and unreliability. Related techniques previously have been developed to fabricate flexible microelectronic and optoelectronic devices on plastic substrates[13, 14, 15], but have not been adapted to silicon integrated photonic circuits[16, 17, 18, 19, 20]. The new method is capable of integrating silicon photonic devices with virtually any technically important substrate materials including sapphire, silicon carbide, diamond, graphene, chalcogenide glasses, and III-V compound semiconductors, just to name a few. As a result, heterogeneously integrated silicon-on-anything photonics can be achieved to potentially realize unprecedented optoelectronic functionalities. As a proof of concept, in this work we demonstrate silicon photonic devices integrated on calcium fluoride ($CaF_2$) substrates to enable low loss operation in the mid-infrared (mid-IR) band and apply this novel platform to demonstrate, for the first time, on-chip cavity-enhanced mid-IR spectroscopic analysis of organic chemicals.

The fabrication process is described here briefly and with more details in the method section. Fig. 1 illustrates the process flow. The starting material is traditional SOI wafers with a top silicon layer bonded on a layer of buried silicon dioxide (BOX). First, an array of holes with a diameter of 4 μm is patterned using photolithography and dry etched into the top silicon layer (Fig. 1a). These holes provide access to the buried oxide layer, which is then partially etched in a hydrogen fluoride (HF) solution. Then the substrate is coated with photoresist. At this time, the photoresist solution is able to go through the holes to encapsulate the underlying, partially etched BOX layer and be cured therein, as shown in Fig. 1b. A flood exposure exposes the photoresist layer except the regions underneath the silicon, which is masked from the exposure of ultra-violet light. After complete removal of the BOX layer using a second, long HF wet etching step, the underlying photoresist structures become pedestals supporting the silicon membrane on top (Fig. 3c). In the next step, a PDMS film is laminated onto the silicon membrane. Because the adhesion force between the photoresist pedestals and the membrane is significantly smaller than that between the PDMS film and the membrane, when the PDMS film is quickly removed from the substrate, the silicon membrane is peeled off with the film (Fig. 3d). To finish the transfer, the PDMS film is pressed on a new receiving substrate and slowly removed by peeling, releasing the silicon membrane bonded on the new substrate by surface forces (Fig. 3e)[21]. Even though it is only the van der Waals and other surface forces that bond the transferred membrane to the substrate, the substrate can be safely rinsed with solvents and coated with photoresist for subsequent patterning with little risk of coming off the membrane. Finally, electron beam lithography and dry etching are used to pattern the transferred silicon membrane on the receiving substrate into strip waveguides, ring resonators and other photonic devices (Fig. 3f).

This method is universally capable of transferring silicon membranes of several square centimeters in area to almost any solid substrates with smooth surfaces. Unlike other methods fro transferring photonic devices[17, 18], our method uses no adhesives so the transferred membrane remains pristine and is free of contamination that may compromise its optical performance and organic adhesive layers that are incompatible with the mid-IR. Using this method, we demonstrate the first silicon photonic devices on crystalline calcium fluoride ($CaF_2$) substrates. $CaF_2$ is highly transparent in the spectral range from visible up to the mid-IR wavelength of 8 μm, and possess a low refractive index of ~1.4 in this wavelength regime. Compared with other mid-IR photonics platforms including silicon-on-sapphire[5, 6] and germanium-on-silicon[22], silicon-on-CaF$_2$ offers both a broad transparency window and the highest index contrast. Fig. 1g shows the image of a finished device successfully fabricated on a CaF$_2$ substrate using the transfer method described above. On the transferred silicon membrane (1.5 cm × 0.8 cm in size), micro-ring resonators coupled with centimeter long rib waveguides arrayed across the membrane are lithographically defined. Because of the high index contrast between silicon and CaF$_2$, the waveguides are very compact with a cross-section of 1.8 μm by 0.6 μm and the micro-ring resonators can have a small bending radius down to 5 μm with low bending loss (<0.1 dB per 90º bend) based on simulation.

Fig. 2 (a-c) shows the optical and scanning electron microscope images of the devices, revealing excellent device quality achieved with the transfer process. The devices' performance in the mid-IR range was tested using a continuous wave, external cavity tunable quantum cascade laser (Daylight Solutions) with a center wavelength of 5.2 μm. Details of the experimental setup and the measurement protocols are elaborated in the method section and the Supplementary Materials. A typical transmission spectrum measured from a device is shown in Fig. 2d. Resonant peaks from two micro-ring resonators with 60 μm radius and nearly critically coupled to the same bus waveguide can be clearly resolved. The optical resonance features an intrinsic quality factor (Q) of $6.2\times10^4$ as shown in Fig. 2e, corresponding to a linear propagation loss of 3.8 dB/cm. Without using resist reflow or post-fabrication etch methods to smooth the waveguide sidewalls, the obtained propagation loss is already among the lowest values in mid-IR silicon waveguides reported so far[5, 6, 23, 24]. Compared with other types of mid-IR silicon photonic devices, our method represents a versatile and robust alternative potentially offering reduced material cost and superior optical performance. Most importantly, the demonstrated capability to transfer crystalline silicon membranes onto arbitrary substrates opens the door to a plethora of exciting possibilities.

The mid-IR spectral range recently has attracted tremendous scientific and technological interests spurred by the rapid strides of mid-IR laser technologies[25]. The mid-IR is also important to spectroscopy and imaging: in particular, mid-IR photonics for absorption spectroscopy promises immense application potentials for chemical and biological sensing in environmental monitoring, homeland security and medical diagnosis. Using the silicon-on-CaF$_2$ devices, we report here the first experimental demonstration of mid-IR on-chip cavity-enhanced spectroscopy and its application to analysis of organic chemicals. Compared to waveguide evanescent sensors[26, 27, 28, 29, 30], this technique can significantly decrease the device footprint while gaining improved sensitivity through capitalizing on the resonantly enhanced, folded optical path length inside the cavity. On-chip cavity-enhanced spectroscopy has previously been applied to chemical and biological analysis in the near-IR telecommunication wavelengths[31, 32, 33, 34, 35]. Transitioning the spectroscopic technology to the mid-IR is expected to dramatically boost its sensitivity, since the characteristic mid-IR absorption bands of chemical or biological species are two to three orders of magnitude stronger than their near-IR overtones. The optical finesse of our mid-IR micro-ring resonators ($F$) is ~190 (corresponding Q-factor $6.2 \times 10^4$) and thus the effective optical path length is $Fn_g R \approx 5$ cm for a micro-ring with radius $R$ of 60 μm, where $n_g = 4.5$ is the group velocity of the fundamental mode. These devices' remarkable optical performance in the mid-IR leads to high sensitivity in our absorption spectroscopy measurement.

In the measurement, the micro-ring resonator was immersed in cyclohexane solutions. Two organic chemicals, ethanol and toluene, were used as the analytes mixed in the solution. Cyclohexane was chosen as the solvent for the spectroscopic analysis given its relatively low optical absorption in the 5.2 μm spectral window. On the other hand, ethanol and toluene have weak absorption peaks centered around 5.2 μm. Fig. 3a and b show a series of transmission spectra of the ring resonator measured in solutions containing different concentrations of ethanol or toluene, respectively. As a result of the excess optical absorption induced by the analytes, we observed that the quality factors and the extinction ratios of the resonance peak decreased systematically with increasing analyte concentration. At the same time, for ethanol the resonant peak blue-shifted with increasing concentration, while for toluene the resonant peak red-shifted. The observed dispersive shift of the resonance frequency is consistent with the dielectric properties of cyclohexane ($n = 1.415$), ethanol ($n = 1.353$) and toluene ($n = 1.475$)[36, 37]. From the measured resonance frequency shift and the extinction ratio change of each resonance peaks in Fig. 2d, the absorption coefficient of the analyte can be calculated at each resonance frequency. Details of the calculation method are presented in the Supplementary Materials. Fig. 3c-d plots the measured absorption spectra of ethanol, toluene and isopropyl alcohol (IPA) in cyclohexane as functions of wavelength using our on-chip spectroscopic interrogation at the discrete resonant wavelengths of the micro-rings. For comparison, results obtained using a traditional benchtop Fourier Transform InfraRed (FTIR) spectrophotometer were also plotted. The excellent agreement between the cavity-enhanced measurement results and the FTIR spectra validates the on-chip spectroscopic sensing mechanism. The small mode volume of the micro-ring cavity further suggests low mass loading limits of detection (LOD) of 0.05 ng for ethanol, 0.06 ng for toluene, and 0.09 ng for IPA, comparable to that of state-of-the-art mid-IR waveguide evanescent sensors[38]. A detailed comparison of the performance metrics of our sensor and previous reports is given in the Supplementary Materials. We note that, because of the limitation of our mid-IR laser, the measurement wavelength in our experiments does not align with the major absorption peak of the tested chemicals. For example, assuming identical device optical performance and using a proper mid-IR laser source, an LOD down to 1 pg is expected simply by shifting the operating wavelength to 3.4 μm, where the absorption of ethanol peaks is 50 times stronger than that at 5.2 μm. The high sensitivity of the cavity-enhanced detection method is attributed to the large field confinement in the sensing region, the small device footprint, as well as the high-Q cavity's resonant enhancement effect.

An additional strength of the demonstrated on-chip cavity-enhanced spectroscopy is its ability to simultaneously determine the changes of both the real and the imaginary part of the complex refractive index ($\tilde{n} = n + i\kappa$) induced by the presence of chemicals in a single spectroscopic scan. The change of the real part is measured from the dispersive shift of the resonance frequency and the change of the imaginary part, related to the absorption coefficient ($\alpha = 4\pi\kappa/\lambda$), is measured from the extinction ratio changes. While the two quantities are fundamentally connected and in principle one can be derived from the measured spectrum of the other using the Kramers-Krönig relation, in practice the measurement spectral range that can be attained with currently available tunable mid-IR lasers is insufficient for such a calculation. The combined index of refraction and optical absorption information can provide spectroscopic fingerprints to achieve unequivocal identification and quantification of chemical species. To illustrate this principle, we performed spectroscopic analysis on two-component mixtures of ethanol and toluene in cyclohexane with varying ratios of concentration, as shown in Fig. 4a. As can be seen in Fig. 3c and d, ethanol and toluene have very different absorption coefficients and refractive indices at 5.2 μm wavelength. Thus, the two parameters can be used as orthogonal parameters in the analysis. In Fig. 4b, the measured absorption coefficient ($\alpha$) and the refractive index change ($\Delta n$) of the mixtures are plotted, along with results from calibration samples containing only ethanol or toluene in cyclohexane. It can be seen that in the $\Delta n - \alpha$ plot, the calibration results from ethanol and toluene of varying concentrations collapse on two straight lines as both $\alpha$ and $\Delta n$ of the solution change proportionally with the concentration. Results from the mixtures fall in between the lines of ethanol and toluene. The observation suggests a simple linear transformation:

$$\begin{bmatrix} c_e \\ c_t \end{bmatrix} = \begin{bmatrix} A & B \\ C & D \end{bmatrix} \begin{bmatrix} \Delta n \\ \alpha \end{bmatrix}$$

which can be performed to directly convert the measured parameter set ($\alpha, \Delta n$) to concentrations ($c_e, c_t$), where $c_e$ and $c_t$ are the concentration of ethanol and toluene, respectively. Thus, the concentrations of the two parts in the mixture can be directly derived from one measurement after the transformation matrix is calibrated (see the Supplementary Materials). The results of chemical concentrations obtained with above method is plotted in Fig. 4c along with values measured using FTIR. It can be seen that the concentrations of the two chemical components agrees nicely with the expected values with errors attributed to calibration inaccuracy. We thus conclude the on-chip cavity-enhanced mid-IR spectroscopy demonstrated here is promising to achieve chemical detection and quantitative analysis for a large variety of chemical species with many important applications.

The results described above demonstrate that our new fabrication method and the achieved silicon-on-anything configuration are promising for mid-IR silicon photonics and can enable novel spectroscopic applications. Without much optimization, the method already yields mid-IR devices with low optical loss comparable to the state-of-the-art, and thus device performance enhancement through further improved processing is anticipated. We also expect that the same approach can be readily applied to fabricate other novel mid-IR silicon photonic devices such as interferometers and photonic crystal cavities on $CaF_2$ substrates. The small mode volume of the photonic crystal cavity, coupled with the strong optical and thermal cavity nonlinearity, will enable ultrasensitive spectroscopic detection and analysis of chemicals in both gaseous and aqueous environments[39, 40].

We believe that this new method we have developed opens door to numerous possibilities of heterogeneous integration of silicon photonics with novel materials. In addition to mid-IR photonics, the method is also generically applicable to silicon photonic integration with other substrate materials with mechanical, electrical or optical properties non-native to silicon and confer unconventional functionalities on silicon photonic circuits. The examples of novel functions that can be potentially enabled by our technology include electro-optic modulation based on nonlinear crystals, flexible photonic integration on plastics, hybrid silicon-graphene photonics, and plasmonic enhancement using metals.

**Method:**

The mid-IR silicon photonic devices were fabricated using silicon-on-insulator (SOI) wafers (Ultrasil Corp.) with 600 nm top silicon layer and 2 μm buried oxide layer. An array of 4 μm diameter holes were subsequently patterned with photolithography and then transferred to the top silicon layer by fluorine based plasma dry etch. A short (~90 seconds) wet etch using concentrated HF (49%) followed to etch through the holes and create slight undercut underneath the rim of each hole and the edge of the membrane. Afterward, Shipley S1813 photoresist was spun coated, flood exposed to ultraviolet light with a mask aligner and developed. The photoresist filled in the undercut regions was masked by the silicon layer, left unexposed, and remained as pedestal support of the silicon membrane after development. The buried oxide layer was then fully removed by another long (~1 hour) wet etch process using concentrated HF (49%). A poly-dimethylsiloxane (PDMS, Corning Sylgard 184) film was fabricated with a mix ratio of 5:1 and cured at 75 ºC for one hour. The PDMS film was laminated onto the SOI substrate, removed quickly to peel off the silicon membrane supported only by the photoresist pedestals. The PDMS film carrying the silicon membrane was then pressed onto a $CaF_2$ substrate and removed slowly to release the membrane and finish the transfer. The photonic structures, including the waveguides and the micro-rings, were then patterned with electron beam lithography (Vistec EBPG 5000+) using ZEP resist and etched into the silicon membrane with fluorine based plasma etch. In the final step, the $CaF_2$ wafer was diced along the edge of the silicon membrane for the fiber end fire coupling.

The organic chemicals used in the spectroscopic tests, ethanol, toluene, isopropyl alcohol, and cyclohexane (> 99.5%), were purchased from Sigma-Aldrich and used as purchased. Cyclohexane was chosen as the blank solvent as it shows low optical absorption around 5.2 μm wavelength. The other three chemicals were used as analytes and their mixtures were prepared based on volume ratios. For the mid-IR on-chip absorption spectroscopy demonstration, the transmission characteristics of the silicon-on-$CaF_2$ micro-rings were carried out on a fiber end fire coupling system, as described in the Supplementary Materials. During the sensing test, the entire surface of the mid-IR resonators sensor chip was covered by drop casted solutions and 16 measurements were performed with each solution composition for statistical averaging. The chip temperature was stabilized at 20 ºC throughout the tests. Due to partial evaporation of the organic components, the solution concentration slowly varied during the optical measurements. Therefore, we used waveguide evanescent wave absorption spectroscopy to calibrate the solution concentration in real time. Transmitted power through a bus waveguide was monitored at the peak absorption wavelengths of the analyte, from which the analyte concentration was calculated based on absorption coefficients of the chemicals measured by FTIR. To independently validate this calibration method, we also used the resonant peak shift and known component refractive indices to extract the analyte concentrations, and both approaches yield identical results. The analyte concentrations evaluated using waveguide evanescent wave absorption spectroscopy were quoted as the nominal concentration values in Figs. 3-4. After optical measurement at each solution concentration, the sensor chip was blown dry immediately to minimize residue formation. In the Fourier transformed infrared spectroscopic measurement, solutions were filled into a demountable liquid cell (PIKE Technologies Inc. 162-1100) with an optical path length varying from 0.1 mm to 0.5 mm and the transmission spectra were recorded by a FTIR spectrometer (Nicolet Magna 860 FTIR).

**Acknowledgements**

This work is supported by the NSF (Award No. CMMI-1200406 and ECCS-1232064). Parts of this work was carried out in the University of Minnesota Nanofabrication Center which receives partial support from NSF through NNIN program, and the Characterization Facility which is a member of the NSF-funded Materials Research Facilities Network via the MRSEC program. H.L. and J.H. also would like to acknowledge technical assistance offered by Y. Chillakuru and K. McLaughlin in setting up the measurement system.

**Author Contributions**

Y.C. and H.L. contributed equally to this work. M.L. and J.H. conceived and designed the experiments. Y.C. developed the transfer process and fabricated the devices. H.L. performed the spectroscopy measurement. All authors contributed to data analysis and co-wrote the paper.

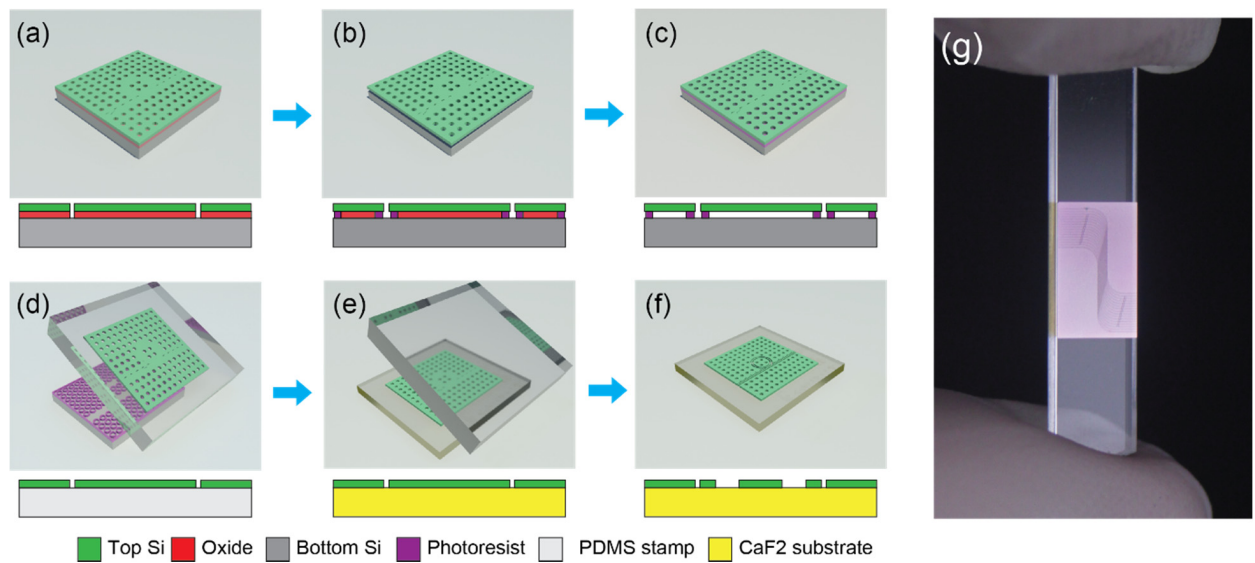

**Figure 1 Fabrication process flow of silicon photonic devices on arbitrary substrates. a-f)** Schematic illustration of the transfer process. **g)** Image of finished silicon photonic devices (light purple area) on a CaF$_2$ substrate (transparent). The silicon membrane has an area of 1.5 cm × 0.8 cm.

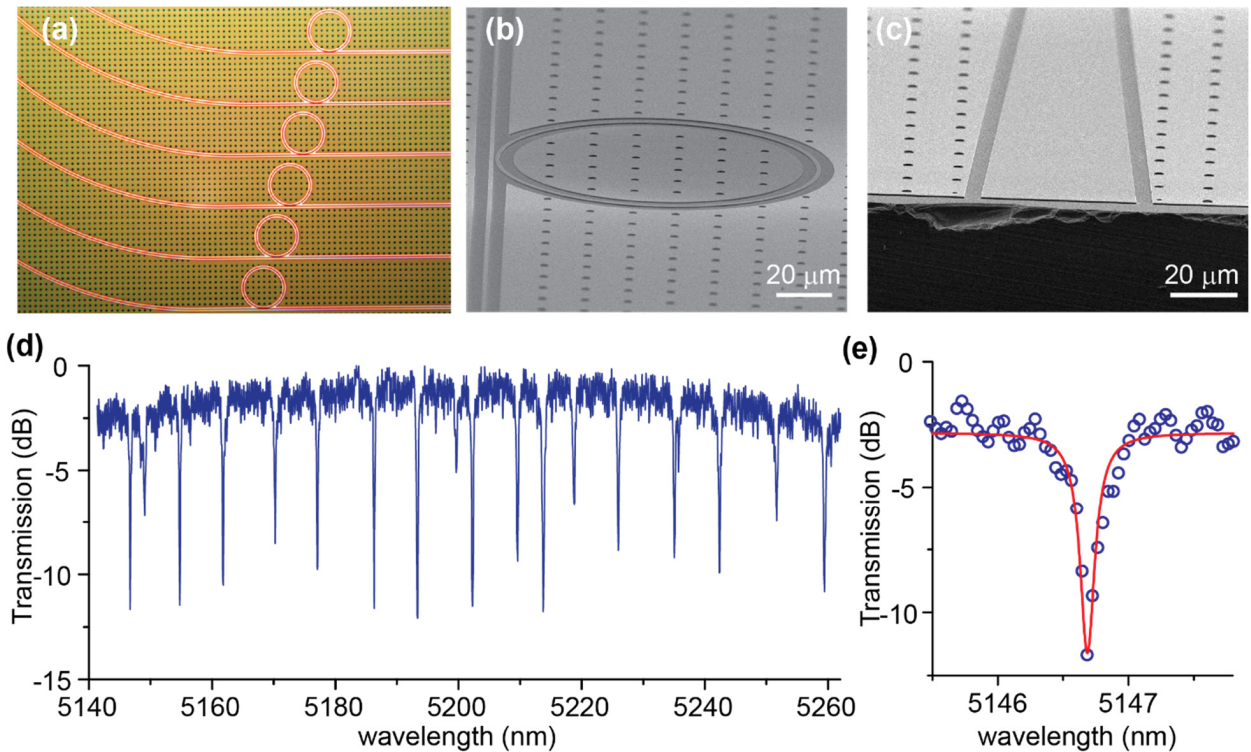

**Figure 2 High performance mid-IR silicon micro-ring resonators on a CaF₂ substrate.** a) Optical and (b, c) scanning electron microscope images of the silicon micro-ring resonators coupled with bus waveguides fabricated on the CaF2 substrate. The waveguides are inverse tapered towards the edge of the chip to facilitate butt-coupling with infrared fibers. (d) The measured transmission spectrum of the device around 5.2 μm, showing resonance peaks from two 60 μm micro-rings. (e) Zoom-in of the resonance: the waveguide loaded Q-factor is $3.7 \times 10^4$, corresponding to an intrinsic Q-factor of $6.2 \times 10^4$ and a propagation loss of 3.8 dB/cm.

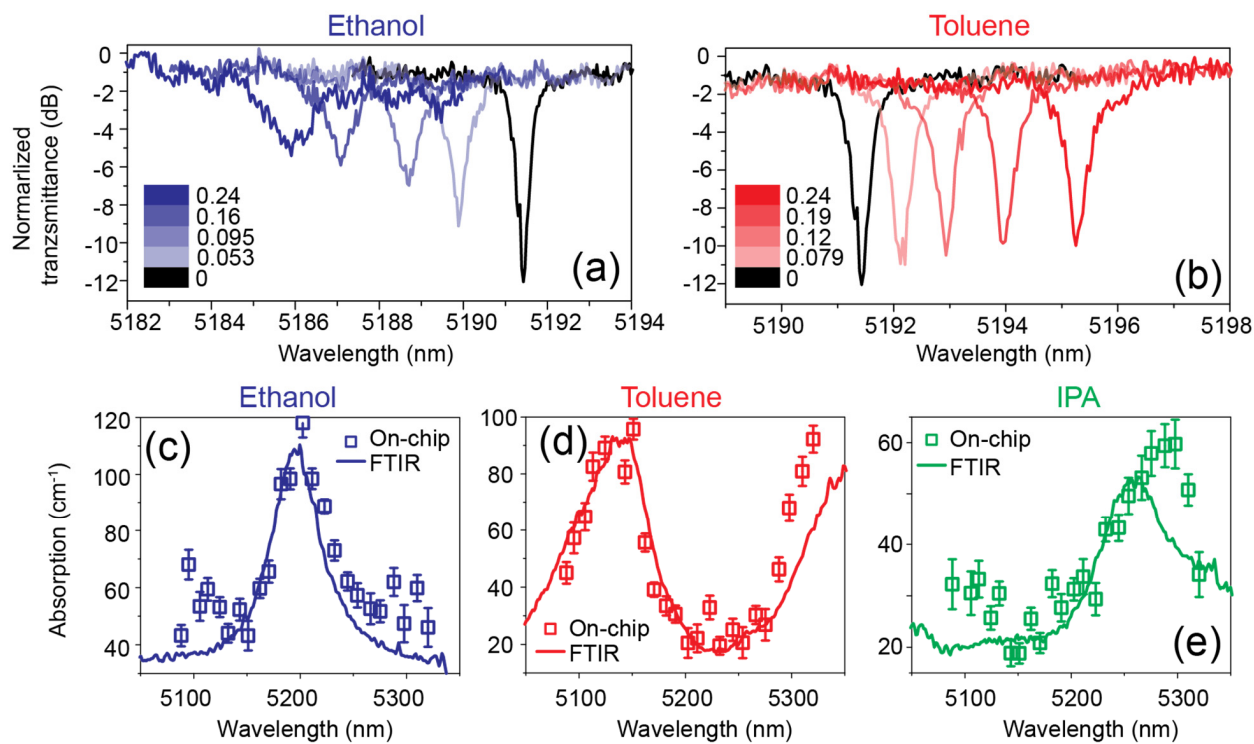

**Figure 3 On-chip mid-IR absorption spectroscopy.** Mid-IR optical transmission spectra of a micro-ring resonator in **(a)** ethanol/cyclohexane and **(b)** toluene/cyclohexane solutions of different nominal concentrations (marked in the legends). Decreasing quality factor and extinction ratio along with frequency shift of the resonance peak can be observed with increasing concentration. **(c-e)** Absorption coefficients of **(c)** ethanol, **(e)** toluene and **(e)** isopropyl alcohol (IPA) measured using the on-chip micro-ring resonators (open squares) and FTIR spectrophotometer (lines), showing excellent agreement between the results obtained by the two methods.

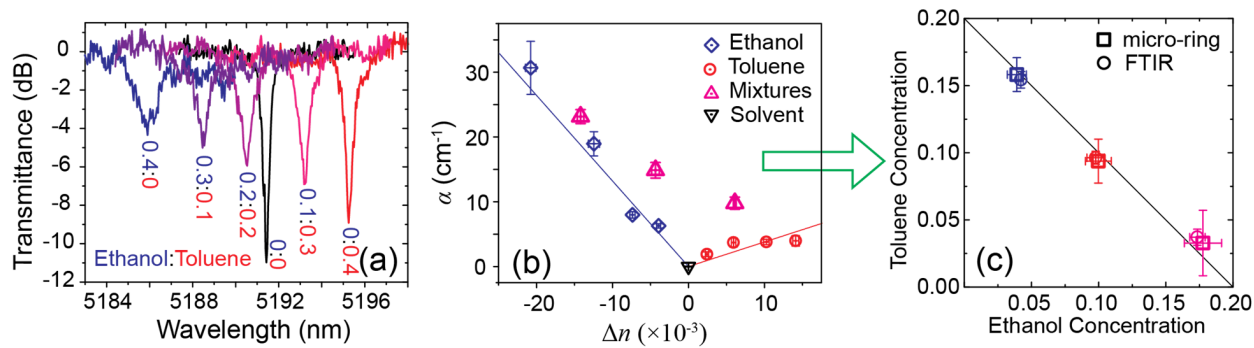

**Figure 4 Chemical analysis of two-component mixture. (a)** Transmission spectra measured from a micro-ring immersed in mixture of ethanol and toluene in cyclohexane with varying concentration ratios as marked on the curves. **(b)** From the measured extinction ratio (ER) and resonance peak shift in (a), the absorption coefficient ($\alpha$) and refractive index change ($\Delta n$) of the mixture (magenta, open triangles) are derived and plotted, along with results from calibration samples (ethanol: blue, open diamonds; toluene: red, open circles; blank solvent: black, open triangle). Both parameters ($\alpha$, $\Delta n$) are linearly dependent on the concentrations and thus can be used to unequivocally determine the concentration of both components. **(d)** After a linear transformation using a matrix determined through calibration, ($\alpha$, $\Delta n$) can be converted to ($c_e$, $c_t$), the concentration of ethanol and toluene in the mixture (open squares), allowing the quantification of their values in a single measurement. Excellent agreement was observed with the nominal values (open circles) obtained by FTIR.